# Relationship between boundary integral equation and radial basis function

Wen CHEN* and Masataka TANAKA**

Department of Mechanical Systems Engineering, Shinshu University, Wakasato 4-17-1, Nagano City, Nagano 380-8553
(E-mail: *chenw@homer.shinshu-u.ac.jp; **dtanaka@gipwc.shinshu-u.ac.jp)

This paper aims to survey our recent work relating to the radial basis function (RBF) from some new views of points. In the first part, we established the RBF on numerical integration analysis based on an intrinsic relationship between the Green's boundary integral representation and RBF. It is found that the kernel function of integral equation is important to create efficient RBF. The fundamental solution RBF (FS-RBF) was presented as a novel strategy constructing operator-dependent RBF. We proposed a conjecture formula featuring the dimension affect on error bound to show the independent-dimension merit of the RBF techniques. We also discussed wavelet RBF, localized RBF schemes, and the influence of node placement on the RBF solution accuracy. The centrosymmetric matrix structure of the RBF interpolation matrix under symmetric node placing is proved.

The second part of this paper is concerned with the boundary knot method (BKM), a new boundary-only, meshless, spectral convergent, integration-free RBF collocation technique. The BKM was tested to the Helmholtz, Laplace, linear and nonlinear convection-diffusion problems. In particular, we introduced the response knot-dependent nonsingular general solution to calculate varying-parameter and nonlinear steady convection-diffusion problems very efficiently. By comparing with the multiple dual reciprocity method, we discussed the completeness issue of the BKM.

Finally, the nonsingular solutions for some known differential operators were given in appendix. Also we expanded the RBF concepts by introducing time-space RBF for transient problems.

## 1. Introduction

The radial basis function (RBF) is an emerging numerical technique for neural network, computational geometry, and more recently, numerical PDEs [1]. It is known that all meshless FEM and BEM schemes based on the moving least square avoid the mesh in matching geometry, however, they still require the mesh to numerical integration [2]. In contrast, the RBF collocation discretization does not require any mesh at all. The method also has no difficulty handling a variety of boundary conditions and is especially promising for high-dimension problems with complex and moving boundary geometry due to the fact that they are independent of dimensionality and inherently meshless.

Despite the excellent performances in some numerical experiments, much reported work concerning RBF solution of PDEs has been mostly based on intuition [3]. In recent years great effort has been devoted to developing a formal mathematical theory of the RBF, however, some very limited advances are achieved by using a sophisticated interpolation analysis [4]. There are still some essential issues regarding the RBF not well studied or even untouched due to great difficulties involved. From some new distinct views of points, recently we have made some significant advances on the RBF mathematical basis and its applications to PDEs [5-7]. The purposes of this paper are to survey our latest work on these challenging issues.

First, we found the intrinsic connection between boundary integral equation and RBF approximation [5,6], which is different from work of Golberg et al. [8,9] in that their research related the RBF with the dual reciprocity method (DRM) rather than directly with boundary integral equation itself. The RBF approach is therefore established here on numerical integration analysis rather than on a sophisticated Hermite-Birkhoff interpolation [4] and native space concepts. The kernel function of integral equation is found the key to construct efficient RBFs. By analogy with Green's second identity, we presented a new fundamental solution RBFs (FS-RBF) construction methodology for inhomogeneous problems in the DRM and the domain-type RBF techniques, which include superconvergent pre-wavelet RBFs [5,6]. Some significant results on RBF construction, error bound and node placement are also provided. The centrosymmetric matrix structure of RBF interpolation matrix is also validated here.

In the second part of this paper, we discussed the boundary knot method (BKM) [5-7]. Kamiya et al. [10] and Chen et al. [11] pointed out that the multiple reciprocity BEM (MRM) solution of the Helmholtz problems with the Laplacian plus its high-order terms is in fact to employ only the singular real part of complete complex fundamental solution of the Helmholtz operator. Power [12] simply indicated that the use of either the real or imaginary part of the Helmholtz Green's representation formula could formulate interior Helmholtz problems. This study extends these ideas to general problems such as Laplace and convection-diffusion problems by a combined use of the nonsingular general solution, DRM, and RBF. This mixed technique is named as the boundary knot method due to its essential meshless property, namely, the BKM does not need any discretization grids for any dimension problems and only uses knot points. Note that the DRM and RBF are employed jointly here to approximate particular solution of inhomogeneous terms as in dual reciprocity BEM (DRBEM) and the method of fundamental solution (MFS). The presented BKM circumvents either singular integration, slow convergence and mesh in the BEM [8] or artificial boundary outside physical domain in the MFS [8,13,14]. Additionally, the present method holds the symmetric matrix structure for self-adjoint operators subject to a kind of boundary condition. The method is integration-free, exponential convergence, meshless, and boundary-only. The term "boundary-only" is used here in the sense as in the DRBEM and MFS that only boundary knots are required, although internal knots can improve solution accuracy. The nonsingular general solution for multidimensional problems can be understood the nonsingular part of complete fundamental solution of various operators.

The preliminary numerical studies show that the BKM is a promising technique in terms of efficiency, accuracy, flexibility, and simplicity. In particularly, we first introduced the response knot-dependent nonsingular general solutions in the BKM calculation of the varying parameter and nonlinear problems very efficiently. The completeness concern of the BKM is also discussed.

This paper is organized as follows. Section 2 establishes the relationship between the boundary integral equation and RBF. The FS-RBF is derived. Error bound, node placement, wavelet RBF and localized RBF are also discussed. In section 3, we introduce the boundary knot method. In particular, we computed varying parameter and nonlinear convection-diffusion problems by the BKM with response knot-dependent nonsingular general solution. Completeness concern of the BKM is discussed. In section 4 numerical results of the BKM are provided and discussed to establish its validity and accuracy. Finally, section 5 concludes some remarks based on the reported results. In appendix, we give the nonsingular general solution of some differential operators. The time-space RBF (TS-RBF) is also presented.

## 2. General solution RBF

In this section, we try to relate the RBF and boundary integral equation through Green's second identity. Without loss of generality, consider Possion equation

$$\nabla^2 u = f(x), x \in \Omega, \quad (1)$$

$$u(x) = D(x), x \subset S_u, \quad (2a)$$

$$\frac{\partial u(x)}{\partial n} = N(x), x \subset S_T, \quad (2b)$$

where $x$ means multi-dimensional independent variable. $n$ is the unit outward normal. The solution of Eq. (1) can be expressed as

$$u = v + u_p, \quad (3)$$

where $v$ and $u_p$ are the general and particular solutions, respectively. The latter satisfies

$$\nabla^2 u_p = f(x) \quad (4)$$

but does not necessarily satisfy boundary conditions. $v$ is the homogeneous solution

$$\nabla^2 v = 0, \quad (5)$$

$$v(x) = D(x) - u_p, \quad (6a)$$

$$\frac{\partial v(x)}{\partial n} = N(x) - \frac{\partial u_p(x)}{\partial n}, \quad (6b)$$

Eqs. (5) and (6a,b) can be solved by various boundary numerical techniques such as the BEM, MFS and BKM.

### 2.1. DRM, RBF and integral equation

In terms of the DRM, the inhomogeneous terms of Eq. (4) are approximated at first by

$$f(x) \cong \sum_{k=1}^{N+L} \alpha_k \phi_k(r_k) \quad (7)$$

where $\alpha_k$ are the unknown coefficients. $N$ and $L$ are respectively the numbers of knots on domain and boundary. $r_k = \|x - x_k\|$ represents the Euclidean distance norm, $\phi$ is the RBF and satisfies

$$\phi(r) = \nabla^2 \psi(r). \quad (8)$$

We can uniquely determine

$$\alpha = A_\phi^{-1} f(x), \quad (9)$$

where $A_\phi$ is nonsingular RBF interpolation matrix. Finally, we can get particular solutions at any point by summing localized approximate particular solutions

$$u_p = \sum_{k=1}^{N+L} \alpha_k \psi_k(r_k). \quad (10)$$

where $\phi$ is also RBF. $u_p$ is evaluated by substituting Eq. (9) into Eq. (10).

Another DRM-equivalent method is to compute the domain integral. By using Green second theorem, the solution of Eq. (1) can be expressed as

$$u(x) = \int_\Omega f(z) u^*(x,z) d\Omega(z) +$$
$$\int_\Gamma \left\{ u \frac{\partial u^*(x,z)}{\partial n(z)} - \frac{\partial u}{\partial n(z)} u^*(x,z) \right\} d\Gamma(z), \quad (11)$$

where $u^*$ is the fundamental solution of Laplace operator. $z$ denotes source point. It is noted that the first and second terms of Eq. (11) respectively equal the particular and general solutions of Eq. (3). If a numerical integral scheme is used to approximate Eq. (11), we have

$$u(x) \cong \sum_{k=1}^{N+L} \omega(x, x_k) f(x_k) u^* +$$
$$\sum_{k=N+1}^{N+L} \omega(x, x_k) \left[ u \frac{\partial u^*}{\partial n} - \frac{\partial u}{\partial n} u^* \right], \quad (12)$$

where $\omega(x, x_j)$ are the weights dependent on the integral schemes. Furthermore, we have

$$u(x) = \sum_{k=1}^{N+L} \alpha_k h_k(x, x_k) u^* f(x_k) +$$
$$\sum_{k=N+1}^{N+L} \beta_k p_k(x, x_k) u^* = \sum_{j=1}^{N+L} \gamma_k \pi_k(x, x_k), \quad (13)$$

where $\alpha$, $\beta$ and $\gamma$ are unknown weighting coefficients, $h_k$ and $p_k$ are the indefinite functions. $hu^* f$, $pu^*$ and $\pi$ can be respectively interpreted as the radial basis functions for the DRM, boundary RBF methods such as the MFS and BKM, and domain-type RBF method. Constructing RBF from numerical integration view of point is named as the fundamental solution RBF [5,6]. How to choose function $h$ and $p$ will be left to discuss later.

RBF $\varphi$ in Eq. (10) can be seen as $hu^*f$ in Eq. (13), which establishes a direct relationship between the RBF and Green integral. If $f(x)$ in Eq. (1) includes some point forcing terms, the RBF approximate representation can be stated as

$$u(x) = \sum_{j=1}^{N+L} \gamma_k \pi_k(x, x_k) + \sum_l^q Q_l u^*(x, x_l), \quad (14)$$

where $q$ is the number of concentrated sources, $Q_l$ is the related magnitude.

## 2.2. Kernel function and RBF

The above analysis show that the RBF has very close relationship with the kernel function of boundary integral equation based on an integral analysis of the RBF solution of PDE. The RBF is now also widely employed in network, data processing, and inverse problems. These problems can be expressed by various integral equations. For instance,

$$Rh = \int_D R(x,y) h(y) dy = f(x), x \in \overline{D} \subset IR^n \quad (15)$$

is the basic equation of stochastic optimization theory, where the $R(x,y)$ is the kernel of a positive rational function of an elliptic self-adjoint operator $L$ on $L^2(IR^n)$ [15]. The creating RBF for this problem should involve the kernel function $R$ as done in Eq. (15).

Another instance is geometry generation which can be understood an elliptic boundary value problems relating to the Laplace and biharmonic operators [16]. The fact that the $\ln(r)$ term in TPS is an essential ingredient of the fundamental solutions of 2D Laplace and biharmonic operator explains why the TPS is widely used in 2D geometry generation problems.

There exists an infinite class of RBFs. Among them, it was experimentally found that the multiquadric (MQ) radial basis function ranked the best in accuracy followed by thin plate spline [17]. The MQ RBF contains a parameter c, which is usually referred to as the shape parameter. The key issue in the MQ in terms of solution accuracy is the proper choice of shape parameter c, which is often problem-independent. So far no general mathematical theory is available to determine its optimal value [3]. The FS-RBF is different from these existing RBFs in that it is created in terms of the operator fundamental solution and inhomogeneous forcing terms. Therefore, the FS-RBF is problem-dependent and generally easy to use. The brutal use of the TPS and MQ should be discouraged. The kernel function is found important to construct efficient operator-dependent RBFs.

## 2.3. Conjecture on the RBF error bounds

The RBF error bounds available now do not include dimension effect. Kansa and Hon [18] observed numerically that the RBF is independent of dimension. They stressed that the convergence order of the RBF increases as the spatial dimension increases, which is featured by error bound $O(M^{(d+1)})$. Here $d$ means the spatial dimension. However, they did not give any evidence supporting this error estimate. Our numerical experiments did not justify this error formula.

It is very interesting to note that there exist the same error behaviors between some RBFs and quasi-Monte Carlo (QMC) or Monte Carlo method (MCM). For example, error bounds for the linear RBF and the classical Monte Carlo method are the same $O(M^{-1/2})$ which $M$ is the number of nodes, while error bounds for the TPS RBF and QMC in 2D problems are the same $O(M^{-1}(\log M))$. It is well known that the QMC has error bound

$$err = O\left(M^{-1}(\log M)^{d-1}\right) \quad (16)$$

for $d$-dimension problems. In fact, the RBF and QMC or MCM have some close relationship on the grounds of numerical integration. The following error bound conjecture of the RBF is intuitively proposed.

**Conjecture 1**: The RBF error bounds can be characterized by $O(M^{-\eta}(\log M)^{d-1})$, where $d$ is dimension, $\eta$ is 1/2 for linear RBF, 1 for the TPS in the 2D problems, and $n$ for the MQ.

The dimension affect on the RBF accuracy is here featured by $O(\log M)^{d-1}$ rather than $O(M^{(d+1)})$ in [18]. The notorious dimension curse in the other numerical techniques can be characterized by

$$err = O(M^{-\kappa/d}), \quad (17)$$

where $\kappa$ denotes the accuracy order of algorithm. By comparing formulas (16) and (17), one can easily conclude that the RBF has very visible advantages in accuracy for higher dimensional problems.

## 2.4. Further analysis of FS-RBF

For inside-domain source points, the FS-RBF creates the RBFs

$$\varphi(x, x_k) = h_k(x, x_k) u^*(x, x_k) f(x_k). \quad (18)$$

Function $h$ assures that the RBF has enough differential continuity when $u^*$ is a singular fundamental solution. Singularity can also be eliminated if we place all response and source points differently or non-singular general solution is used [5,7]. In such cases, $h$ can be understood a RBF weight function. $h=r^{2m}$ is a convenient choose where $r$ is the Euclidean distance. One can find that the generalized thin plate spline function $r^{2m}\ln(r)$ is a special case of the FS-RBF for the Laplace and biharmonic operators in 2D problems. $r^{2N+1}$ should be recommended for higher-dimensional problems.

Although Duchon [19] showed that the TPS with linear constraints is optimal for biharmonic operator interpolation, the linear constraints are not considered necessary if we modify TPS to $r^2\ln(r)+r^2+1$ according to the FS-RBF approach, where $\ln(r)$ general solution of biharmonic operator is not used due to its singularity. As of the boundary source points, we suggest

$$\varphi(x) = p_k(x, x_k) u^*(x, x_k) \quad (19)$$

as the RBF. $p$ functions like $h$ in Eq. (18).

What follows is an in-depth analysis of $h$ and $p$. The method of reduction of variance which includes importance sampling and stratified sampling are important to improve the accuracy of the MCM. Consider the integrand

$$\int w d\Omega = \int (w/g) g d\Omega = \int R g d\Omega, \quad (20)$$

where $gd\Omega$ denotes density which means the variable transformation method using the indefinite integral of $g$. The basic idea is to try to reduce function $R$ as close as possible to constant so that the accuracy of the MCM integral is considerably increased. This suggests that *h* and *p* should be positive definition with purpose averaging integrand function. Gaussian RBF may be interpreted as Gaussian probability distributions which are cases in some practical situations. Some sophisticated MCM techniques can be effectively used to construct efficient RBF within this framework. Also various singularity removal techniques in the BEM can be employed to construct RBF and MCM.

One immediate question is which of the MCM algorithm is a counterpart of the known MQ with spectral accuracy. For Green integral (11), we have

$$\int f(x)u^*(r,x)d\Omega = \int \sqrt{r^2+c^2}\frac{f(x)u^*(r,x)}{\sqrt{r^2+c^2}}d\Omega \qquad (21)$$
$$= \int \sqrt{r^2+c^2}g(r,x,c)d\Omega,$$

Comparing with Eq. (20), the shape parameter *c* of the MQ is here interpreted as giving the random points different weights corresponding to their position relative to other points in the MCM. Gamma distribution density may be the closest to the MQ. It is conceived that if the quasi-random nodes are weighted through varying parameter *c* to keep the MQ constant as possible, the MCM algorithm can be the spectral convergence. However, this scheme may not be feasible due to the difficulty in indefinite integral of *g*. In contrast, the flexibility of the RBF lies in that it always can enforce arbitrary variable transformation in the reduction of variable method via collocation technique. This finding give some explanations to Kansa' variable shape parameter MQ [18].

The RBF may be applicable to evaluate any high-dimensional integral. Let integrand function in Eq. (20)

$$f(x) = w(x)/u^*(s_i,x) \qquad (22)$$

where $s_i$ is one specified node. Integrand (20) can be transformed to the evaluation of particular solution of Possion equation (1) with *f(x)* defined in Eq. (22) as forcing term and zero Dirichlet boundary condition. Seeming inefficiency of this strategy due to the solution of a PDE can be eased if we use the spectral accuracy MQ as the RBF in the DRM. In addition, the multipole, multigrid and wavelet techniques can be used to reduce effort in the inversion of the RBF interpolation matrix to O(*M*log(*M*)) [20].

On the other hand, $hu^*f$ (or $pu^*$) from integral weight function viewpoint should be orthogonal with respect to node distribution just as in multidimensional orthogonal wavelet series. For simplicity, considering an ideal case where integrand function can be expressed only in terms of radial variable, we can reduce it to 1D problem

$$\int_a^b h(r)u^*(r)f(r)dr = \sum_{k=1}^M v_k h(r_k)u^*(r_k)f(r_k), \qquad (23)$$

where $v_k$ is weight coefficients. $hu^*$ is seen as the weight function in Gauss integration which can be chosen to remove integral singularities and make integral smooth. In the RBF case, the distribution of variable *r* changes with response point. In this case we have abscissas at first and then we need to determine the weight function with a recurrence procedure to maximize the degree of accuracy with radical orthogonality. This is more or less similar to Gauss-Kronrod quadrature.

On the other hand, the system matrix of the globally-supported RBF methods is prone to severe ill-conditioning. It is noted that one source point in the RBF fashion is restricted within a certain region where it can impose significant affect on the other points. In fact, the collocation scheme can be understood as the maximum order finite difference method (FDM) of global support. Therefore, it is practically attainable to truncate the global support into finite support RBF in a way like the FDM [6]. Another strategy localizing RBF is to use domain decomposition with or without overlapping knots between inside-domain boundaries. Consequently, we have a banded sparse system matrix of resulting discretization equations. Such localized RBF schemes are called as the finite knot method (FKM).

### 2.5. Nodes placement

Fedoseyev et al [21] found that compared with uniform nodes, the use of non-uniform ones are often much more efficient in the domain RBF solution of PDE. Furthermore, Fedoseyev et al. [22] addressed the severe accuracy drop of RBF PDE solutions at nodes neighboring boundary. Powell [20l] also found that the RBF geometry interpolation encounters evident accuracy loss around boundary. In fact, such edge effects are well known in polynomial interpolation and differential approximation, which are due to the geometry and system equation discontinuities at boundary. Remedy for geometry discontinuity is to place the nodes inclining boundary such as the use of zeros spacing of the Lobatto, Chebyshev or legendre polynomials [23]. This methodology also works for the system discontinuity problem. Therefore, the domain RBF scheme for PDE should use non-uniform nodes. A quantitative measure for node distributions is

$$\sigma(x_i) = \frac{\Omega}{M}\sum_{k=1}^M \|x_i - x_k\|. \qquad (24)$$

Geometry discontinuity can be understood that $\sigma$ value of distinct node is different, especially between boundary and central nodes. Fedoseyev et al. [22] seem to solve the system discontinuity problem very well by collocating the governing equation at boundary points.

### 2.6. Coefficient matrix structures and ill-conditioning

It is well known that the RBF interpolation matrix has symmetric structure irrespective of the geometry and node placements. This study will show that the RBF matrix also carries the centrosymmetric structure if the nodes are symmetrically placed. Let us consider a *d*-dimension space problem. The Euclidean distance norm is defined as

$$r_{ij} = \sqrt{\sum_{k=1}^d \left(x_i^{(k)} - x_j^{(k)}\right)^2}. \qquad (25)$$

The symmetric node spacing is understood as

$$x_i^{(k)} + x_{N+1-i}^{(k)} = x_j^{(k)} + x_{N+1-j}^{(k)} = c^{(k)}, \qquad (26)$$

where $c^{(k)}$ are constants. Thus, it is obvious

$$r_{ij} = r_{N+1-i,N+1-j}. \qquad (27)$$

Based on Eq. (27), one can easily verify that the RBF coefficient matrices have centrosymmetric structure as shown in Eq. (27) for even order derivative and skew centrosymmetric structure as shown below

$$a_{ij} = -a_{N+1-i,N+1-j} \qquad (28)$$

for odd order derivative, where *a* denotes an entry of a matrix. Therefore, if the nodes are symmetrically placed, the RBF coefficient matrices for derivatives have either symmetric centrosymmetric or skew symmetric centrosymmetric structures.

Centrosymmetric matrices can easily be decomposed into two half-sized matrices. Such factorization merit leads to a considerable reduction in computing effort for determinant, inversion and eigenvalues. For more related details see [24,25]. On the other hand, even if the total node spacing is not symmetric, such decomposition processing can still effectively reduce the ill-conditioning of the resulting RBF matrices by preconditioning

$$\hat{A} = \begin{bmatrix} I & J \\ 0 & I \end{bmatrix} A \begin{bmatrix} I & 0 \\ -J & I \end{bmatrix} \quad (29)$$

where $A$ means a RBF matrix of even order, $J$ is contra-identity matrix. A similar but distinct preconditioning transform matrix exists for odd order matrix.

In addition, we observe that as in the traditional collocation method [26], very large entries of RBF derivative matrices appear in the upper and lower two rows and middle columns and largely account for the ill-conditioning of large-size RBF system. Accordingly some elementary matrix transformations can simply significantly reduce the RBF ill-conditioning with very little effort.

**2.7. Wavelet RBF**

Chen and Tanaka [5,6] found that if we replace distance variable $r$ by $\sqrt{r^2 + c^2}$ in the FS-RBF, we got the prewavelet RBF with the exponential accuracy, where $c_j$ is dilution parameters. For example, numerical experiments with pre-wavelet TPS $r_j^{2m} \ln \sqrt{r_j^2 + c_j^2}$ manifests spectral convergence as in the MQ. This work can be generalized by

$$u = \sum_{k=1}^{N} c_k \phi(\lambda_k r_k + d_k), \quad (30)$$

where $\lambda$ and $d$ are respectively dilate and location coefficients of the wavelet. $\phi$ is the RBF which can be here seen as a wavelet parent function. Such wavelet-like RBF may be especially attractive for adaptable handling geometry singularity and localized shock-like solutions due to its inherent multiscale feature combined with a spatial localization.

Fasshauer and Schumaker [27] summarized some wavelets using sphere RBFs. It will be beneficial to pay more attentions on this aspect, especially for creating orthonormal wavelet-like RBF.

**3. Boundary knot method**

One may think that the placement of source points outside domain in the MFS is to avoid the singularities of fundamental solutions. However, we found through numerical experiments that even if all source and response points were placed differently on physical boundary to circumvent the singularities, the MFS solutions were still degraded severely. In the MFS, the more distant the source points are located from physical boundary, the more accurate MFS solutions are obtained [8]. However, unfortunately the resulting equations can become extremely ill-conditioned which in some cases deteriorates the solution [8,13,14].

Like the DRBEM and MFS, the BKM can be viewed as a two-step numerical scheme, namely, DRM and RBF approximation to particular solution and the evaluation of homogeneous solution. The latter is the emphasis of this paper. The former has been well developed now [8]. We outline the basic methodology to approximate particular solution in the previous section 2.1. For more details see [5-8]. What follows is procedure calculating homogeneous solution in the BKM.

**3.1 Nonsingular general solution and BKM**

To illustrate the basic idea of the boundary collocation using nonsingular general solution, we take the 2D Helmholtz operator as an illustrative example, which is the simplest among various often-encountered operators having nonsingular general solution. The Laplace operator has not nonsingular general solution. For the other nonsingular general solutions see appendix.

The 2D homogeneous Helmholtz equation (8) has two general solutions, namely,

$$v(r) = c_1 J_0(r) + c_2 Y_0(r), \quad (31)$$

where $J_0(r)$ and $Y_0(r)$ are the zero-order Bessel functions of the first and second kinds, respectively. In the standard BEM and MFS, the Hankel function

$$H(r) = J_0(r) + i Y_0(r) \quad (32)$$

is applied as the fundamental solution. It is noted that $Y_0(r)$ encounters logarithm singularity, which causes the major difficulty in applying the BEM and MFS. Many special techniques have been developed to solve or circumvent this singular trouble.

The present BKM scheme discards the singular general solution $Y_0(r)$ and only use $J_0(r)$ as the radial function to collocate the boundary condition Eqs. (6a) and (6b). It is noted that $J_0(r)$ exactly satisfies the Helmholtz equation and we can therefore get a boundary-only collocation scheme. Unlike the MFS, all collocation knots are placed only on physical boundary and can be used as either source or response points.

Let $\{x_k\}_{k=1}^{N}$ denote a set of nodes on the physical boundary, the homogeneous solution $v(x)$ of Eq. (5) is approximated in a standard collocation fashion

$$v(x) = \sum_{k=1}^{N} \beta_k J_0(r_k), \quad (33)$$

where $r_k = \|x - x_k\|$. $k$ is the index of source points. $N$ is the number of boundary knots. $\beta_k$ are the desired coefficients. Collocating Eqs. (6a) and (6b) in terms of Eq. (33), we have

$$\sum_{k=1}^{N} \beta_k J_0(r_{ik}) = b_1(x_i) - u_p(x_i), \quad (34a)$$

$$\sum_{k=1}^{N} \beta_k \frac{\partial J_0(r_{jk})}{\partial n} = b_2(x_j) - \frac{\partial u_p(x_j)}{\partial n}, \quad (34b)$$

where $i$ and $j$ indicate Dirichlet and Neumann boundary response knots, respectively. If internal nodes are used, we need to constitute another set of supplement equations

$$\sum_{k=1}^{N}\beta_k J_0(r_{lk}) = u_l - u_p(x_l), \quad l=1,\ldots,L, \quad (35)$$

where $l$ indicates the internal response knots and $L$ is the number of interior points. Now we get total $N+L$ simultaneous algebraic equations. It is stressed that the use of interior points is not always necessary in the BKM, although internal knots can improve solution accuracy in some cases as in the DRBEM and MFS [28,29].

### 3.2. Completeness concern

One of major potential concerns of the BKM is its completeness due to the fact that the BKM employs only the non-singular part of fundamental solutions of differential operators. This incompleteness may limit its utility. Although the given numerical experiments favor the method, now we can not theoretically ascertain of the general applicability of the BKM. On the other hand, Kamiya and Andoh [10] validated that the similar incompleteness occurs in the multiple reciprocity BEM using the Laplacian for Helmholtz operators. Namely, if the Laplace fundamental solution plus its higher-order terms are used in the MRM for the Helmholtz problems as in its usual form, we actually employ only the singular part of Helmholtz operator fundamental solution. Although the MRM performed well in many numerical experiments, it is mathematically incomplete. It is interesting to note that the BKM and MRM respectively employ the nonsingular and singular parts of the complete complex fundamental solution. It should also be stressed that although Power [12] simply indicated that the singular or nonsingular parts of Green representation can formulate the interior Helmholtz problems, no any related numerical and theoretical results are available from the published reports.

Kamiya and Andoh [10] also pointed out that the MRM formulation with the Laplacian could not satisfy the well-known Sommerfeld radiation conditions at infinity. Chen et al. [11] addressed the issues relating to spurious eigenvalues of the MRM with the Laplacian. Some literatures referred to in [11] also discussed the issues applying MRM with the Laplacian to problems with degenerate boundary conditions. These concerns of the MRM raise some cares concerning the applicability of the BKM which implements the nonsingular part of fundamental solution compared to the MRM using the singular part. Power [12] discussed the incompleteness issue of the MRM for the Brinkman equation and indicated that using one part of complex fundamental solution of Helmholtz operator may fail to the exterior Helmholtz problems. However, now we can not justify whether or not the BKM works for exterior problems since the method differs the MRM in using the DRM approximation of particular solutions. Dai [30] successfully applied the dual reciprocity BEM with the Laplacian to waves propagating problems in an infinite or semi-infinite region. It is worth pointing out that the Laplace fundamental solution used in the DRBEM also does not satisfy the Sommerfeld radiation condition. Unlike the MRM, the BKM and DRBEM do not employ the higher-order fundamental solutions to approximate the particular solution. Our next work will investigate if the BKM with the DRM can analyze the unbounded domain problems.

In fact, all existing numerical techniques encounter some limits. The BKM is not exceptional. Power [12] pointed out that the incompleteness in the MRM is problem-dependent. Therefore, the essential issue relating to the concerns of the BKM completeness is under what conditions the method works reliably and efficiently. In addition, there exist some controversies in the choice of basis functions or fundamental solution even in so-called established technique. For example,

the biharmonic operator has four general solutions

$$w^*(r) = C_1 \ln(r) + C_2 r^2 \ln(r) + C_3 r^2 + C_4. \quad (36)$$

However, it is common practice in the BEM to only use $r^2 \ln(r)$ as the fundamental solution. We are wondering if the absence of the other three terms in the BEM fundamental solution may cause some completeness concerns in some situations. Although Duchon [19] proved that $r^2 \ln(r)$ is optimal interpolants for biharmonic operator with linear polynomial constraints, the only use of $r^2 \ln(r)$ is mathematically incomplete. Also Duchon's linear polynomial constraints are artificial conditions.

## 4. Numerical results and discussions

In this paper, all numerical examples unless otherwise specified are taken from [28]. The geometry of test problem is all an ellipse featured with semi-major axis of length 2 and semi-minor axis of length 1. These examples are chosen since their analytical and numerical solutions are obtainable to compare. More complicated problems can be handled in the same BKM fashion without any extra difficulty. The MQ is employed as the RBF in terms of the DRM. The zero order Bessel and modified Bessel functions of the first kind are evaluated via short subroutines given in [31]. The 2D Cartesian co-ordinates (x,y) system is used as in [28].

### 4.1. Helmholtz equation

Consider the inhomogeneous 2D Helmholtz problem governed by equation

$$\nabla^2 u + u = x. \quad (37)$$

Inhomogeneous boundary condition

$$u = \sin x + x \quad (38)$$

is posed. It is obvious that Eq. (38) is also a particular solution of this problem. Numerical results by the present BKM is displayed in Table 1.

**Table 1. Results for the Helmholtz problem**

| x | Y | Exact | BKM (5) | BKM (7) |
|---|---|---|---|---|
| 1.5 | 0.0 | 2.50 | 2.45 | 2.51 |
| 1.2 | -0.35 | 2.13 | 2.08 | 2.14 |
| 0.6 | -0.45 | 1.16 | 1.18 | 1.16 |
| 0.0 | 0.0 | 0.0 | 0.1 | -0.002 |
| 0.9 | 0.0 | 1.68 | 1.66 | 1.69 |
| 0.3 | 0.0 | 0.60 | 0.64 | 0.60 |
| 0.0 | 0.0 | 0.0 | 0.08 | -0.001 |

The numbers in brackets of Table 1 mean the total nodes. The shape parameter $c$ in is chosen 3. It is found that the present BKM converges very quickly. This shows the BKM holds the super-convergent merit. The BKM can yield accurate solutions with only 7 knots. In contrast, the DRBEM employs 16 boundary and 17 interior points to achieve slightly less accurate solutions for a simpler homogeneous case [28].

This is because the normal BEM has only low order of convergence speed [8], while the BKM keeps the spectral convergence of the collocation-type techniques [23].

### 4.2. Laplace equation

This case will further justify the superconvergence of the BKM through a comparison with the BEM for Laplace equation

$$\nabla^2 u = 0 \qquad (39)$$

with boundary condition

$$u = x + y. \qquad (40)$$

Eq. (40) is easily found to be a particular solution of the problem. This homogeneous problem is typically well suited to be handled by the standard BEM technique. In contrast, there is inhomogeneous term in the BKM formulation to apply the nonsingular general solution of the Helmholtz operator. Namely, Eq. (39) is rewritten as

$$\nabla^2 u + u = u, \qquad (41)$$

where the right inhomogeneous term $u$ is approximated by the DRM as shown in the section 2. The numerical results are displayed in Table 2 where the BEM solutions come from [28].

The MQ shape parameter $c$ is set 25 for both 3 and 5 boundary knots in the BKM. It is observed that the BKM solutions are not sensitive to the parameter $c$. It is seen from Table 2 that the BKM results using 3 boundary nodes achieve the accuracy of four significant digits and are far more accurate than the BEM solution using 16 boundary nodes. This striking accuracy of the BKM again validates its spectral convergence. In this case only boundary points are employed to approximate the particular solution by the DRM and RBF. It is noted that the coefficient matrices of the BEM and BKM are both fully populated. Unlike the BEM, however, the BKM yields symmetric coefficient matrix for all self-adjoint operators with one type of boundary conditions. This Laplace problem is a persuasive example to verify high accuracy and efficiency of the BKM vis-a-vis the BEM.

**Table 2. Results for the Laplace problem**

| X | y | Exact | BEM (16) | BKM (3) | BKM (5) |
|---|---|---|---|---|---|
| 1.5 | 0.0 | 1.500 | 1.507 | 1.500 | 1.500 |
| 1.2 | -0.35 | 0.850 | 0.857 | 0.850 | 0.850 |
| 0.6 | -0.45 | 0.150 | 0.154 | 0.150 | 0.150 |
| 0.0 | 0.0 | -0.450 | -0.451 | -0.450 | -0.450 |
| 0.9 | 0.0 | 0.900 | 0.913 | 0.900 | 0.900 |
| 0.3 | 0.0 | 0.300 | 0.304 | 0.300 | 0.300 |
| 0.0 | 0.0 | 0.0 | 0.0 | 0.0 | 0.0 |

### 4.3. Convection-diffusion problems

The FEM and FDM encounter some difficulty to produce accurate solution to the systems involving the first order derivative of convection term. Special care need be taken to handle this problem with these methods. It is claimed that the BEM does not suffer similar accuracy problem. In particular, the DRBEM was said to be very suitable for this type problem [28,32]. Let us consider the convection diffusion equation

$$\nabla^2 u = -\partial u / \partial x, \qquad (42)$$

which is given in [28] to test the DRBEM. The boundary condition is stated as

$$u = e^{-x}, \qquad (43)$$

which also constitutes a particular solution of this problem. By adding $u$ on dual sides of Eq. (42), we have

$$\nabla^2 u + u = u - \partial u / \partial x. \qquad (44)$$

The results by both the DRBEM and BKM are listed in Table 3.

**Table 3. Results for** $\nabla^2 u = -\partial u / \partial x$

| x | y | Exact | DRBEM (33) | BKM (15) | BKM (18) |
|---|---|---|---|---|---|
| 1.5 | 0.0 | 0.223 | 0.229 | 0.229 | 0.224 |
| 1.2 | -0.35 | 0.301 | 0.307 | 0301 | 0.305 |
| 0.0 | -0.45 | 1.000 | 1.003 | 1.010 | 1.000 |
| -0.6 | -0.45 | 1.822 | 1.819 | 1.822 | 1.818 |
| -1.5 | 0.0 | 4.482 | 4.489 | 4.484 | 4.477 |
| 0.3 | 0.0 | 0.741 | 0.745 | 0.744 | 0.743 |
| -0.3 | 0.0 | 1.350 | 1.348 | 1.353 | 1.354 |
| 0.0 | 0.0 | 1.000 | 1.002 | 1.003 | 1.004 |

**Table 4. Results for** $\nabla^2 u = -\partial u / \partial x - \partial u / \partial y$

| x | y | Exact | DRBEM (33) | BKM (15) | BKM (18) |
|---|---|---|---|---|---|
| 1.5 | 0.0 | 1.223 | 1.231 | 1.225 | 1.224 |
| 1.2 | -0.35 | 1.720 | 1.714 | 1.725 | 1.723 |
| 0.0 | -0.45 | 2.568 | 2.557 | 2.546 | 2.551 |
| -0.6 | -0.45 | 3.390 | 3.378 | 3.403 | 3.405 |
| -1.5 | 0.0 | 5.482 | 5.485 | 5.490 | 5.491 |
| 0.3 | 0.0 | 1.741 | 1.731 | 1.729 | 1.731 |
| -0.3 | 0.0 | 2.350 | 2.335 | 2.349 | 2.350 |
| 0.0 | 0.0 | 2.000 | 1.989 | 1.992 | 1.993 |

The MQ shape parameter is chosen 4. The BKM employed 7 boundary knots and 8 or 11 internal knots. In contrast, the DRBEM [28] used 16 boundary and 17 inner nodes. It is stressed that unlike the previous examples, in this case the use of the interior points can improve the solution accuracy evidently. This is due to the fact that the governing equation has convection domain-dominant solution. Only by using boundary nodes, the present BKM with the Helmholtz non-singular solution and the DRBEM with the Laplacian [28] can not well capture convection effects of the system equation. It is found from Table 3 that both the BKM and DRBEM achieve the salient accurate solutions with inner nodes. The BKM outperforms the DRBEM in computational efficiency due to the super-convergent features of the MQ interpolation and global BKM collocation.

Further consider equation

$$\nabla^2 u = -\partial u/\partial x - \partial u/\partial y \qquad (45)$$

with boundary conditions

$$u = e^{-x} + e^{-y}, \qquad (46)$$

which is also a particular solution. The numerical results are summarized in Table 4.

In the BKM, the MQ shape parameter is taken 5.5. We employed 7 boundary knots and 8 or 11 inner points in the BKM compared with 16 boundary nodes and 17 inner points in the DRBEM [28]. The BKM worked equally well in this case as in the previous ones. It is seen from Table 4 that the BKM with fewer points produced almost the same accurate solutions as the DRBEM. Considering the extremely mathematical simplicity and easy-to-use advantages of the BKM, the method is superior to the DRBEM in this problem.

### 4.4. BKM with response knot-dependent nonsingular general solution

It is very difficult to find the fundamental solutions for various linear varying parameter and nonlinear differential operators. In linear varying parameter cases, a normal strategy [32] is to decompose variable parameters into two parts, an average constant and a perturbation, where the constant is included into the fundamental solution of the specified operator. In nonlinear case, an iterative linearization procedure is usually employed [28]. Similarly, it is a daunting task to constitute nonsingular general solutions for these problems. In this section, we use the response knot-dependent nonsingular general solution to significantly simplify the BKM solution of such problems. It is noted that this strategy can be extended to the BEM.

#### 4.4.1. Varying-parameter Helmholtz problem

Consider the varying-parameter Helmholtz equation

$$\nabla^2 u - \frac{2}{x^2} u = 0 \qquad (47)$$

with inhomogeneous boundary condition

$$u = -2/x. \qquad (48)$$

Eq. (48) is also a particular solution of this problem. Note that the origin of the Cartesian co-ordinates system is dislocated to the node (3,0) to circumvent singularity at $x=0$. The response knot-dependent nonsingular general solution of varying parameter equation (47) is given by

$$u(r_{ik}, x_i) = I_0\left(\frac{\sqrt{2}}{|x_i|} r_{ik}\right), \qquad (49)$$

where $I_0$ is the zero order modified Bessel function of the first kind. $i$ and $k$ respectively index the response and source nodes. In terms of the BKM, the problem can be analogized by

$$\sum_{k=1}^{N} \alpha_k I_0\left(r_{ik}\sqrt{2}/x_i\right) = -2/x_i. \qquad (50)$$

Note that only the boundary nodes are used in Eq. (50). After evaluating the coefficients $\alpha$, we can easily evaluate the value of $u$ at any inner node $p$ by

$$u_p = \sum_{k=1}^{N} \alpha_k I_0\left(r_{pk}\sqrt{2}/x_p\right). \qquad (51)$$

Table 5 lists the BKM results against the DRBEM solutions. The BKM average relative errors under $N=9$, 13, 15 are respectively 9.7e-3, 8.1e-3 and 7.6e-3, which numerically demonstrates its convergence. The accuracies of the BKM and DRBEM solutions are comparable. It is noted that the DRBEM used 33 nodes (16 inner and 17 boundary knots) in this case, while the BKM only employed much less boundary knots. The accuracy and efficiency of this BKM scheme are very encouraging. Note that the present BKM representation differs from the previous ones in that we here use the response point-dependent nonsingular general functions.

**Table 5. Relative errors for varying parameter Helmholtz problem**

| x | y | DRBEM(33) | BKM (9) | BKM (15) |
|---|---|---|---|---|
| 4.5 | 0.0 | 2.3e-3 | 3.3e-3 | 2.6e-3 |
| 4.2 | -0.35 | 2.1e-3 | 4.1e-3 | 3.3e-3 |
| 3.6 | -0.45 | 5.4e-3 | 6.8e-3 | 4.7e-3 |
| 3.0 | -0.45 | 4.5e-3 | 1.1e-2 | 4.4e-3 |
| 2.4 | -0.45 | 1.2e-3 | 1.4e-2 | 9.1e-4 |
| 1.8 | -0.35 | 9.0e-4 | 5.2e-3 | 1.7e-2 |
| 3.9 | 0.0 | 3.9e-3 | 7.0e-3 | 5.3e-3 |
| 3.3 | 0.0 | 3.3e-3 | 1.1e-2 | 6.3e-3 |
| 3.0 | 0.0 | 4.5e-3 | 1.3e-2 | 5.6e-3 |
| 2.7 | 0.0 | 2.7e-3 | 1.5e-2 | 3.4e-3 |
| 2.1 | 0.0 | 3.2e-3 | 1.6e-2 | 8.8e-3 |

#### 4.4.2. Nonlinear convection-diffusion problems

Chen and Tanaka [5,6] found that if only boundary knots are used, the BKM can formulate linear analogization of nonlinear differential equations with linear boundary conditions. Consider the Burger-like convection-diffusion equation

$$\nabla^2 u - u_x u = 0 \qquad (52)$$

with inhomogeneous boundary condition

$$u = -2/x \qquad (53)$$

Note that the previous varying parameter problem is in fact the simplified version of this nonlinear case. Eq. (53) is also a particular solution. Using the scheme given in [5], Eq. (52) is restated as

$$\nabla^2 u + u = u + u_x u. \qquad (54)$$

When only the boundary nodes are employed, the resulting BKM formulation will be a linear algebraic equation in terms of 2D Helmholtz non-singular general solution. For details of such BKM procedure see [5]. The numerical solutions of this normal BKM procedure have the average relative errors 3.9e-2 for $N=5$, 1.1e-1 for $N=9$, 1.4e-1 for $N=13$ at some inner nodes, where $N$ is the number of boundary nodes used. It is noted that the performances are unstable and solutions inaccurate. If we use interior points in the BKM, the accuracy and stability will be improved greatly at the expense of sacrificing linear formulation as in the dual reciprocity BEM [28]. It is noted that the convection term rather than nonlinear constitution here causes the deficiency of the BKM solutions if not using inner points.

In the DRBEM, it is reported [28] that the use of the fundamental solution of convection diffusion equation can significantly improve the solution accuracy of transient convection diffusion problem with only one inner point. This suggests us that a non-singular general solution of convection-diffusion operator may be much more suitable for this problem. By analogy with the fundamental solution of the convection-diffusion equation, the response knot-dependent non-singular general solution of Eq. (52) is given by

$$u(r,x) = e^{-u(x-x_k)/2} \phi\left(\frac{|u|}{\sqrt{2}} r\right), \quad (55)$$

where radial function $\phi$ is the zero order Bessel $J_o$ or modified Bessel function $I_o$ of the first kind dependent on the sign of flow velocity $u$. In this case it is the latter. The RBF approximation is given by

$$u_i = \sum_{k=1}^{N} \alpha_k e^{-u_i(x_i-x_k)/2} I_0\left(\frac{|u_i|}{\sqrt{2}} r_{ik}\right). \quad (56)$$

Note that the above RBF representation differs from the normal one in that we here use the response point-dependent RBFs even if we do not know the value of $u$. In terms of the BKM using only boundary knots, Eq. (52) is analogized by substituting boundary conditions into Eq. (56), namely,

$$\sum_{k=1}^{N} \alpha_k e^{(x_i-x_k)/x_i} I_0\left(r_{ik}\sqrt{2}/x_i\right) = -2/x_i. \quad (57)$$

The formulation (57) is a set of simultaneous linear algebraic equations and can be solved easily by any linear solver.

Then we can evaluate the value of $u$ at any inner nodes of interest through the solution of a single nonlinear equation (56). Note that the RBF expansion coefficients in Eq. (56) are now already evaluated from Eqs. (57) and the only one unknown is the value of $u$ at a specified single inner knot. This study used simple bisection method to handle such a single nonlinear equation. There are no concerns here relating to the expensive repeated evaluation and inverse of Jacobian matrix, stability issue and the careful guess of initial solutions.

In summary, the present nonlinear BKM scheme can be viewed as a two-step procedure. First, the linear BKM formulation of nonlinear problems is produced using response node-dependent RBFs. Then, the second step is to calculate the solution at any inner node of interest through solving a single nonlinear algebraic equation.

Table 6 lists the BKM results compared with the solutions of the dual reciprocity BEM [28]. Average relative errors of the present BKM for $N=9, 11$ are respectively 8.5e-3, 7.5e-3. Compared with the previous BKM procedure, the solution accuracy is significantly improved while still keeping the linear BKM formulation of nonlinear problems. However, it is noted that the solution accuracy is still not always improved with incremental number of boundary knots. For example, average relative errors for $N=13, 15, 17, 19, 21$ are respectively 8.3e-3, 8.3e-3, 8.8e-3, 8.9e-3, 1.9e-2. Namely, the highest average accuracy is achieved for N=11. As in the other global collocation techniques, this is due to the ill-conditioning system matrix for large number of knots [23]. Anyway overall solution procedure is rather stable. The accuracy and efficiency of the present BKM scheme are very encouraging.

**Table 6. Relative errors for BKM linear formulation of Burger-like equation**

| x | y | DRBEM(33) | BKM (9) | BKM (11) |
|---|---|---|---|---|
| 4.5 | 0.0 | 2.3e-3 | 2.8e-3 | 2.5e-3 |
| 4.2 | -0.35 | 2.1e-3 | 2.3e-3 | 2.9e-3 |
| 3.6 | -0.45 | 5.4e-3 | 4.4e-3 | 6.2e-3 |
| 3.0 | -0.45 | 4.5e-3 | 1.0e-2 | 9.2e-3 |
| 2.4 | -0.45 | 1.2e-3 | 1.2e-2 | 5.7e-3 |
| 1.8 | -0.35 | 9.0e-4 | 7.0e-3 | 3.2e-3 |
| 3.9 | 0.0 | 3.9e-3 | 4.1e-3 | 5.5e-3 |
| 3.3 | 0.0 | 3.3e-3 | 9.1e-3 | 1.0e-2 |
| 3.0 | 0.0 | 4.5e-3 | 1.2e-2 | 1.1e-2 |
| 2.7 | 0.0 | 2.7e-3 | 1.4e-2 | 1.1e-2 |
| 2.1 | 0.0 | 3.2e-3 | 1.1e-2 | 4.9e-3 |

On the other hand, the solutions of the DRBEM [28] were obtained with 16 boundary nodes and 17 inner points. Therefore, it is not surprising that the DRBEM solutions [28] are slightly more accurate than the present BKM ones. It is also noted that the DRBEM formulation is a set of simultaneous nonlinear algebraic equations. The programming, computing effort and storage requirements in the DRBEM are much higher than the BKM. The present case only involves the Dirichlet conditions. For more complicated boundary conditions, the resulting BKM formulation may not be a set of linear algebraic equations even if we only employ boundary knots. By any measure, however, the size of the BKM analogous equations will be much smaller than that of the DRBEM.

Similarly, we can easily constitute response node-dependent fundamental solutions. Thus, the essential idea behind this work may be extended to the BEM, DRBEM and MRM for varying parameter linear and nonlinear problems. Namely, as in handling linear varying parameter problems, the response knot-dependent fundamental solution may be employed in the nonlinear computing even if some parameters of fundamental solution are unknown. For example, unlike the DRBEM scheme for varying velocity convection-diffusion problems given in Ref. [32], the variable convection-diffusion fundamental solutions with response node-dependent velocity parameters may be employed to the BEM or the DRBEM formulations.

## 5. Concluding remarks

Golberg and Chen [8] put the DRM on the firm RBF theory. The present work further establishes direct relationship between the RBF itself and the integral equation. The presented FS-RBF fully exploits the features of certain problems and is applicable to the domain-type RBF and DRM approximations of various inhomogeneous problems. The proposed FS-RBF, TS-RBF, and prewavelet RBFs may be fundamentally important to the RBF theory and its broad applications to PDEs, neural network, computational geometry, data processing and inverse problems. A comparative study of these new RBFs and popular MQ, TPS, compactly-supported RBFs

[33] is under way.

The present BKM can be regarded one kind of the Trefftz method [34] where the trial function is required to satisfy governing equation. The BKM distinguishes from the other Trefftz techniques such as the MFS in that we employ nonsingular general solution. The shortcomings of the MFS using fictitious boundary are eliminated in the BKM. The term "BKM" can be interpreted as a boundary modeling technique combining the DRM, RBF, and nonsingular general solution.

To preserve symmetric matrix structure for self-adjoint operator problems with mixed conditions, the BKM boundary collocation can be locally performed within the same type of boundary condition separately. These collocation equations in different local boundary zones will then be matched through keeping $C_0$ continuity at interface knots. The symmetric matrix is known well-conditioned [35].

In conclusion, the presented BKM is inherently possesses some desirable numerical merits which include meshless, boundary-only, integration-free, exponential convergence, and mathematical simplicity. The implementation of the method is remarkably easy. More numerical experiments to test the BKM will be beneficial.

## Appendix

By its very basis, it is straightforward to extend the BKM to the nonlinear, three-dimensional, time-dependent partial differential systems. The following lists the nonsingular general solutions of some important steady and transient differential operators.

For the 3D Helmholtz-like operators

$$\nabla^2 u \pm \lambda^2 u = 0, \qquad (A1)$$

we respectively have the nonsingular general solution

$$u^* = A \frac{\sin(\lambda r)}{r} \qquad (A2)$$

and

$$u^* = A \frac{\sinh(\lambda r)}{r}, \qquad (A3)$$

where sinh denotes the hyperbolic function, $A$ is constant. For the 2D biharmonic operator

$$\nabla^4 w - \lambda^2 w = 0, \qquad (A4)$$

we have the nonsingular general solution

$$w^* = A_1 J_0(\lambda r) + A_2 I_0(\lambda r). \qquad (A5)$$

The non-singular general solution of the 3D biharmonic operator is given by

$$w^* = A_1 \frac{\sin(\lambda r)}{r} + A_2 \frac{\sinh(\lambda r)}{r}. \qquad (A6)$$

For the 3D time-dependent heat and diffusion equation

$$\Delta u = \frac{1}{k} \frac{\partial u}{\partial t}, \qquad (A7)$$

we have the nonsingular general solution

$$u^*(r,t,t_k) = A e^{-k(t-t_k)} \frac{\sin(r)}{r}. \qquad (A8)$$

Furthermore, consider 3D transient wave equation

$$\Delta u = \frac{1}{c^2} \frac{\partial^2 u}{\partial t^2}, \qquad (A9)$$

we get the general solution

$$u^*(r,t,t_k) = \left[ A_1 \cos(c(t-t_k)) + \frac{A_2}{c} \sin(c(t-t_k)) \right] \frac{\sin(r)}{r}. \qquad (A10)$$

The 2D nonsingular general solutions of transient problems can be easily derived in a similar fashion.

Two standard techniques handling time derivatives are the time-stepping integrators and the model analysis. The former involves some difficult issues relating to the stability and accuracy, while the latter is not very applicable for many cases such as shock. The BKM using time-dependent nonsingular solutions may circumvent these drawbacks. The difficulty implementing such BKM schemes may lie in how to satisfy the inharmonic initial conditions inside domain as in the time-dependent BEM. On the other hand, the time-dependent nonsingular general solutions may be directly applied in the domain-type RBF collocation schemes such as the Kansa's method. It is worth pointing out that the analogous method proposed by Katsikadelis et al. [29] may be combined with the BKM to handle the differential systems which do not include Laplace or biharmonic operators.

On the other hand, the time-space RBF [6] is presented for domain RBF collocation methods. The time-space distance function is given by

$$r_j = \sqrt{(x-x_j)^2 + c(t-t_j)^2}, \qquad (A11)$$

which can be applied to the wave propagation problem. Here $c$ is the wave velocity. The definition (A11) of distance function differs from the standard radial distance function in that the time variable is handled equally as the space variables. The RBFs with such time-space distance variable eliminates time dependence easily.

For parabolic-type diffusion equation, we can define the time-space RBF as

$$\phi(r_p,t,t_j) = h(r_p,t,t_j) u^*(r,t,t_j), \qquad (A12)$$

where transient fundamental solution $u^*$ is

$$u^* = \frac{1}{(t_j-t)^{d/2}} \exp\left(-\frac{r_p^2}{4k(t_j-t)}\right) H(t_j-t). \qquad (A13)$$

$d$ is the space dimensionality, $H$ is the Heaviside function. $h$ is chosen according to problem feature. It is stressed that in this case the response and source nodes must be totally staggered to avoid singularity in time dimension.

## Acknowledgements:

The authors express grateful acknowledgement of helpful discussions with C.S. Chen, M. Golberg. Y.C. Hon, J.H, He and A.H.D. Cheng. The first author is supported as a JSPS postdoctoral research fellow by the Japan Society of Promotion of Science.